\documentclass[9pt,twocolumn]{extarticle}
\PassOptionsToPackage{table}{xcolor}
\usepackage{classicarxiv}
\usepackage{algorithmic}
\usepackage{makecell}
\usepackage[linesnumbered,ruled,vlined]{algorithm2e}
\usepackage{multirow}
\usepackage{xspace}
\usepackage{pifont}
\usepackage{tabularx}
\usepackage{siunitx}
\usepackage[normalem]{ulem}

\newcommand{\bn}[1]{}
\newcommand{\zc}[1]{}
\newcommand{\re}[1]{#1}
\newcommand{\codename}{\texttt{AMARA}\xspace}
\newcommand{\ir}{MCMC-IR\xspace}

\renewcommand{\boldsymbol}[1]{\symbf{#1}}

\title{AMARA: Compiler-Driven Accelerator Architecture Design for Graphical Probabilistic Sampling}
\author[1,2]{Zhan Chen}
\author[1,2]{Yihan Fu}
\author[3]{Wenshuo Yue}
\author[1]{Daijing Shi}
\author[2]{Chenyun Yin}
\author[1,2]{Mingrui Wu}
\author[4]{Xiangyu Zhang}
\author[2]{Yun Liang}
\author[1,2,*]{Bonan Yan}
\affil[1]{Institute for Artificial Intelligence, Peking University, China}
\affil[2]{Beijing Advanced Innovation Center for Integrated Circuits, School of Integrated Circuits, Peking University, China}
\affil[3]{School of Computing, National University of Singapore, Singapore}
\affil[4]{Independent Researcher}
\affil[*]{Corresponding author: \href{mailto:bonanyan@pku.edu.cn}{bonanyan@pku.edu.cn}}
\date{}

\shorttitle{AMARA: Compiler-Driven Accelerator Architecture Design}
\shortauthor{Zhan Chen and Coauthors}
\hypersetup{
  pdftitle={AMARA: Compiler-Driven Accelerator Architecture Design for Graphical Probabilistic Sampling},
  pdfauthor={Zhan Chen, Yihan Fu, Wenshuo Yue, Daijing Shi, Chenyun Yin, Mingrui Wu, Xiangyu Zhang, Yun Liang, Bonan Yan}
}

\begin{document}

\firstpageornament[26mm]{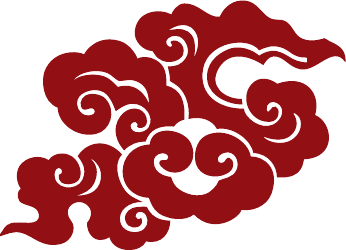}

\makeatletter
\twocolumn[%
\begin{@twocolumnfalse}
  \maketitle
\begin{abstract}

  Sampling methods are fundamental primitives for training and inference on probabilistic models.
  However, practical deployment of sampling algorithms such as Markov Chain Monte Carlo (MCMC) remains difficult on general-purpose processors. Existing hardware accelerators improve sampling efficiency, but they are typically specialized for a narrow class of models, graph topologies, or update schedules, making it difficult to support the diverse workloads used in real applications.
  This paper presents \codename{}, an end-to-end compiler-driven accelerator framework for probabilistic sampling on arbitrary probabilistic graphical models.
  \codename{} introduces \ir, a domain-specific intermediate representation that separates local factorized computation from global update dependencies. Based on this representation, the compiler performs architecture-aware partitioning, scheduling, and memory-access optimization to expose parallelism and reduce irregular graph communication. The optimized program is lowered to a compact ISA and executed on a programmable ASIC accelerator with specialized compute, sampling, and memory support.
  By co-designing the compiler and the architecture, \codename{} bridges the gap between model generality and accelerator efficiency for practical MCMC-based probabilistic computing.
  We evaluate \codename{} across representative workloads with different structures, including grid-structured and irregular energy-based models, and causal inference networks.
  The results show that \codename{} supports a broad spectrum of discrete probabilistic models while achieving high sampling throughput and efficient hardware utilization.
  \codename{} achieves an average of $6.1\times$ (up to $16.3\times$) higher sampling throughput than a JAX-GPU baseline on irregular workloads, and offers broader programmability than state-of-the-art accelerators.

\end{abstract}
\end{@twocolumnfalse}
]
\makeatother

\begingroup
  \renewcommand{\thefootnote}{}
  \footnote{This paper has been accepted to HPCA 2027.}
  \addtocounter{footnote}{-1}
\endgroup

\section{Introduction}
\label{sec:intro}

Probabilistic computing represents and manipulates uncertainty through probability distributions, stochastic state transitions, and sampling, supporting applications such as structured perception, Bayesian and causal inference, and combinatorial optimization~\cite{zhu1998filters,hosseini2020bayesian,mohseni2022ising}.
Probabilistic graphical models (PGMs) express a joint distribution through local interactions among random variables~\cite{koller2009probabilistic}, but exact inference for large-scale PGMs is often intractable, making Markov Chain Monte Carlo (MCMC) sampling an important primitive~\cite{barbu2020introduction}.
This work targets component-wise sampling algorithms such as Gibbs sampling~\cite{geman1984stochastic} over discrete PGMs.

\begin{figure}[t!]
\centering
\includegraphics[width=\linewidth]{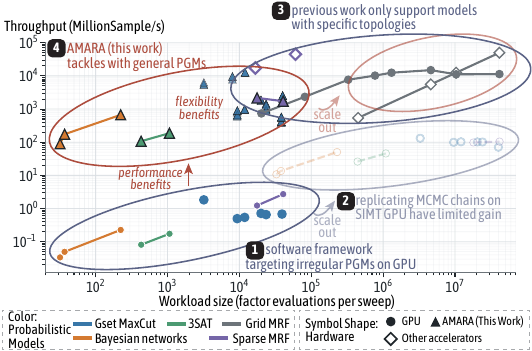}
\caption{Comparison of sampling performance and scaling properties for different workloads on different computing platforms.}
\label{fig:intro-challenge}
\end{figure}

Despite their broad applicability, MCMC samplers remain difficult to accelerate across model families.
A sampler is specified through local probabilistic interactions and update dependencies but must be translated into coordinated computation, data movement, and parallel execution.
This mapping is straightforward only for regular model families with predictable structure; for general PGMs, heterogeneous kernels and instance-specific topologies require compilation before execution.
As shown in Fig.~\ref{fig:intro-challenge}, attainable throughput depends jointly on workload size and structure and on the platform's mapping efficiency.

Existing platforms address only part of this mapping problem.
Illustrated in Fig.~\ref{fig:intro-challenge}, CPUs and GPUs preserve flexibility, but their common remedy of replicating independent Markov chains improves aggregate throughput without shortening the convergence path of an individual chain.
\re{Chain-replication does not scale linearly as well, increasing the number of parallel chains by 4096$\times$ yields only about a 270$\times$ throughput gain.} 
Specialized accelerators~\cite{bybee2023efficient,iyer2025efficient,garg2025efficient,aadit2022massively,ko2020pgma,zhang2018optical,zhang2021robustness,chai2022coopmc,kjellqvist2025biglittle,zhao2024aia,zhao2025aia,zhao2025mc2a,banerjee2019acmc2} improve efficiency but often assume a fixed topology, factor family, or update schedule.
Consequently, existing accelerator solutions occupy only limited regions of the workload scale and structure in Fig.~\ref{fig:intro-challenge}, rather than providing a single scalable execution path for diverse PGMs.

The challenge is not merely to build another accelerator for a particular probabilistic model.
Programmable accelerators require a compiler to expose legal parallelism and locality; otherwise, large irregular workloads remain difficult to map and scale efficiently.
\re{If unoptimized, simply increasing the parallel resources of a programmable accelerator brings limited speedup for irregular workloads. Increasing the processing unit count by $16\times$ reduces sweep cycles by only $1.53\times$.} 
\label{position:evidence}
These observations motivate a compiler-driven co-design that translates model semantics into architecture-compatible computation and communication.

We present \codename{}, a compiler-driven ASIC framework for MCMC sampling over diverse discrete PGMs, co-designing an execution-oriented IR, architecture-aware compilation, and a programmable sampling accelerator to make model semantics executable as efficient static hardware programs.
Our contributions are:
\begin{enumerate}
  \item \textbf{Compiler-visible PGM updates.} \ir{} separates the local factorized computation of each update kernel from inter-kernel update constraints, enabling independent lowering of local computation and global scheduling.
  \item \textbf{Efficient static mappings for irregular PGMs.} The compiler lowers local computations to compact ISA templates and applies partitioning and coloring to reduce cross-block communication and load imbalance.
  \item \textbf{Hardware-friendly programmable execution.} The \codename{} architecture executes compiler-generated MCMC programs for diverse discrete PGM workloads.
\end{enumerate}

Evaluated across grid-structured and irregular energy-based models and causal inference networks, \codename{} delivers an average of $6.1\times$ (up to $16.3\times$) higher sampling throughput than a JAX-GPU baseline on irregular workloads while supporting a broader spectrum of discrete probabilistic models than state-of-the-art accelerators.
More broadly, the \ir{} abstraction and backend optimization flow make model-specific mapping decisions explicit for probabilistic-computing accelerators. \codename is open-sourced at \url{https://anonymous.4open.science/r/AMARA}.

\section{Preliminaries}
\subsection{Probabilistic Graphical Models}
\label{sec:preliminary}

\begin{figure}[t!]
\centering
\includegraphics[width=\linewidth]{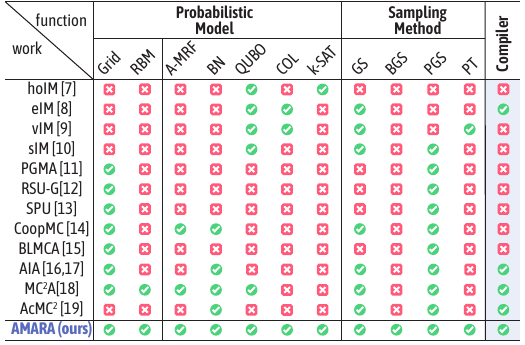}
\caption{Qualitative comparison of supported model classes and programmability in prior accelerators.
}
\label{fig:table1}
\end{figure}

A probabilistic graphical model defines a joint probability distribution $P (\boldsymbol{V})$ over a set of nodes representing random variables $\boldsymbol{V} = \{v_1, v_2, \ldots, v_n\}$.
Each variable $v_i \in \{1, 2, \ldots, L_i\}$ takes one of $L_i$ categorical states.
Dependencies (in Equation~\ref{eq:dep}) and interactions (in Equation~\ref{eq:interact}) are represented by edges in the graph.
\begin{align}
    P(\boldsymbol{V}) &= \prod_{i=1}^{n} P (v_i | \mathrm{Parent}(v_i)) \label{eq:dep}\\
    P(\boldsymbol{V}) &= \frac{1}{Z} \prod_{c \in \mathcal{C}} \psi_c (\boldsymbol{V}_c) \label{eq:interact}
\end{align}

A common representation to describe probabilistic graphical models is \textbf{factor graph}~\cite{opengm-benchmark,opengm-library,DGM,KosovPhdThesis,jelincic2025efficient,zhou2022pgmax,banerjee2019acmc2,tensorlogic_quantrs_hooks,loeliger2004introduction}, i.e., a bipartite graph $\{V, F, E\}$ containing random variables $v\in V$ and factor nodes $f \in F$ that represent relationships among variables. A factor can be evaluated by inputting the neighboring variables $\boldsymbol{V}_{\mathrm{clique}(f)}$ to a function $\phi_f$, then the joint distribution can be factorized as the product of all factors.
\label{position:factor-graph}
\begin{align}
   P(\boldsymbol{V}) &\propto \prod_{f\in F}\phi_f (\boldsymbol{V}_{\mathrm{clique}(f)}) \\
  \boldsymbol{V}_{\mathrm{clique}(f)} &= \{
    v | (v,f)\in E
  \}
\end{align}

\paragraph*{Processing Challenge~--~Topology}
Probabilistic graphical models can be classified by whether their topology follows a structural template or is defined by an irregular problem instance. This distinction is important for hardware acceleration because template-structured models often admit predetermined memory layouts, dataflows, and parallel schedules, whereas instance-structured models require topology-aware compilation to derive efficient hardware mappings.

\textbf{Template-structured models} have topology patterns that are compactly described by a model family, such as lattice-like or bipartite interaction structures. This regularity enables predictable memory access patterns and reusable parallel schedules. 
Examples include the restricted Boltzmann machine (RBM)~\cite{hinton2012practical,fischer2012introduction}, which adopts a bipartite structure, 
and the Grid Markov Random Field (Grid MRF), widely used in unsupervised computer vision tasks~\cite{barnard1989stochastic,wilson2003class,szeliski2008comparative}. 
\textbf{Instance-structured models}, in contrast, carry topology as part of each input problem instance. They often exhibit heterogeneous connectivity, making efficient mapping dependent on per-instance scheduling, placement, and memory optimization.
Examples include MRFs on arbitrary graphs (e.g.,~\cite{nowozin2011decision,niazi2024training,dattani2019pegasus,boothby2021zephyr}) to model more complex relationships, Bayesian networks (BN) for causal reasoning~\cite{kitson2023survey,hosseini2020bayesian,arora2019bayesian}, and combinatorial optimization problems (COP), ranging from quadratic unconstrained binary optimization (QUBO) or Ising Model~\cite{lucas2014ising,mohseni2022ising} that only uses binary states and pairwise interaction, to multi-valued (e.g., graph coloring~\cite{johnson1991optimization,kawakami2023constrained,garg2025efficient} and travel salesman problem~\cite{matai2010traveling,dan2020clustering}) and higher-order (e.g., $k$-SAT~\cite{schoning1999probabilistic,bybee2023efficient}) formulations.

\uline{Currently, few DSAs support the full spectrum of these graphical models (Fig.~\ref{fig:table1}).} Moreover, aside from a few customized compilers~\cite{banerjee2019acmc2,iyer2025efficient,zhao2025aia,zhao2025mc2a}, a unified compilation framework capable of mapping irregular graphical models onto diverse DSAs is urgently needed, especially as many new DSAs continue to emerge.

\subsection{Sampling Methods}

\begin{figure}[t!]
\centering
\includegraphics[width=\linewidth]{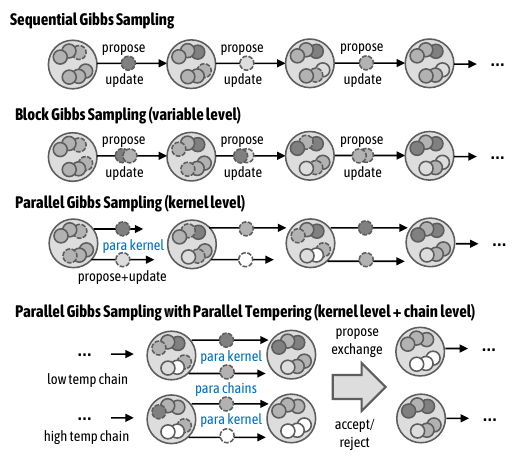}
\caption{Gibbs Sampling variants with different parallelism.}
\label{fig:Gibbs-Variant}
\end{figure}

For expressive or large-scale probabilistic models with intractable energy functions or high-dimensional spaces, exact inference becomes infeasible, necessitating approximate inference methods such as MCMC sampling.~\cite{koller2009probabilistic,barbu2020introduction}
For constraint satisfaction problems and optimization problems, sampling and annealing methods also provide a unified framework to map and solve these problems.~\cite{kirkpatrick1983optimization,mohseni2022ising}

MCMC algorithms sample from the model distribution by evolving a Markov chain $\{\Omega, \nu, K\}$, where $\Omega$ denotes the state space, $\nu$ is the initial distribution, and $K$ is the transition kernel.
The transition kernel is constructed such that the Markov chain can converge to the target distribution $\pi$. 
Gibbs sampling (GS)~\cite{geman1984stochastic} is a widely used MCMC algorithm for probabilistic graphical models, whose transition kernel can be decomposed into a sequence of subkernels that use the full conditional distribution of the target distribution.
\begin{align}
  K_{\mathrm{GS}}(x\to x') &= \prod_{i=1}^{|V|} K_{i,\mathrm{GS}}(x_{i-1}\to x_i) \label{eq:GS} \\
  K_{i,GS}(x_{i-1}\to x_i) &= P(x_i|x_{-i}) \label{eq:GS_k}
\end{align}



\paragraph*{Processing Challenge~--~Parallelization}
Fig.~\ref{fig:Gibbs-Variant} illustrates several Gibbs sampling variants that accelerate the convergence of the Markov chain using different forms of parallelism.
Block Gibbs sampling (BGS)~\cite{titsias2017hamming} updates a group of variables jointly, increasing the update granularity and exposing intra-kernel computational parallelism in evaluating the block conditional distribution.
Parallel Gibbs sampling (PGS)~\cite{gonzalez2011parallel} exploits kernel-level parallelism by updating conditionally independent variables simultaneously, subject to dependency constraints among proposals.
Parallel tempering (PT)~\cite{earl2005parallel,mingas2012parallel} further introduces multiple replicas at different temperatures, enabling chain-level parallelism and improving exploration of multimodal state spaces.

However, these variants are often treated as standalone algorithmic optimizations. 
Existing accelerators mainly rely on variable-level parallelism, and typically support only fixed parallel update patterns like chessboard coloring scheduling for the grid topology model.
This limits their ability to uniformly map diverse sampling schemes onto hardware, especially when the irregular model topologies bring more parallelization opportunities.

\section{Compiler-Driven Design Framework}
\label{sec:flow}

\codename{} uses a compiler-driven stack that makes model semantics explicit before mapping them to a programmable accelerator.
Model topology and the sampling algorithm are abstracted hierarchically as computation kernels and their dependencies, so that the compiler can derive both executable instructions and architecture-aware schedules for component-wise MCMC sampling.

\begin{figure}[t!]
\centering
\includegraphics[width=\linewidth]{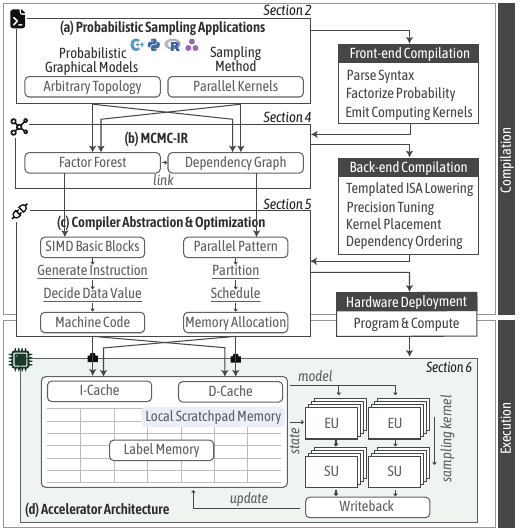}
\caption{\codename framework: the proposed compiler-driven lowering (software-to-hardware) stack.
}
\label{fig:framework}
\end{figure}

Fig.~\ref{fig:framework} illustrates the end-to-end \codename{} workflow.
At compile time, a PGM specification is translated into \ir, optimized by the backend compiler, and lowered into ISA instructions together with schedules and configuration metadata.
At runtime, the host loads this compiled program and the initial label states onto the accelerator; the accelerator then iteratively evaluates update kernels, samples and updates new labels.
For models with the same topology but different parameters or inputs, the compiled instructions and schedules can be reused by updating only data or initial labels; models with different topologies are supported through recompilation.

\paragraph{MCMC Intermediate Representation (\ir)} Fig.~\ref{fig:framework}(b) is the compiler front-end representation for component-wise MCMC over discrete PGMs.
It consists of two linked abstractions: a \emph{Factor Forest}, which describes the local factorized computation of each update kernel, and a \emph{Dependency Graph}, which captures mutual-exclusion constraints among update kernels for sequential iteration.
The Factor Forest determines what each update kernel computes, while the Dependency Graph determines which kernels can be placed and scheduled together.
Sec.~\ref{sec:ir} details these two abstractions.

\paragraph{Backend lowering and optimization} Fig.~\ref{fig:framework}(c) converts \ir into an architecture-aware representation.
The backend lowers Factor Forest trees into compact ISA basic blocks through instruction templates. Relevant data values are converted to the fixed-point format used in the hardware.
A static kernel placement and an order phase schedule are derived from the independent-set parallel pattern of the Dependency graph.
The resulting schedule preserves dependency constraints while seeking to reduce cross-block label communication and phase-level workload imbalance for irregular PGMs.
Sec.~\ref{sec:backend} describes this backend flow. 

\paragraph{\codename Accelerator Architecture} Fig.~\ref{fig:framework}(d) depicts the programmable ASIC architecture used to execute compiled \codename{} programs.
Energy Units (EUs) evaluate candidate-label log-likelihoods from compiled factor instructions, cached model data, and label states; Sampler Units (SUs) transform these values into sampled labels.
The memory system supplies both model information and mutable state: instruction caches store compiled instructions, data caches store model parameters, and hierarchical label memories provide high-bandwidth label access while preserving visibility for labels shared across blocks.
This organization represents graph-specific connectivity, scheduling, and state placement through compiler-generated instructions and metadata, rather than hardwiring a fixed graph topology into the datapath.
The hardware architecture is described in Sec.~\ref{sec:arc}.
\section{Front-End: MCMC Intermediate Representation}
\label{sec:ir}

\begin{figure}[t]
\centering
\includegraphics[width=\linewidth]{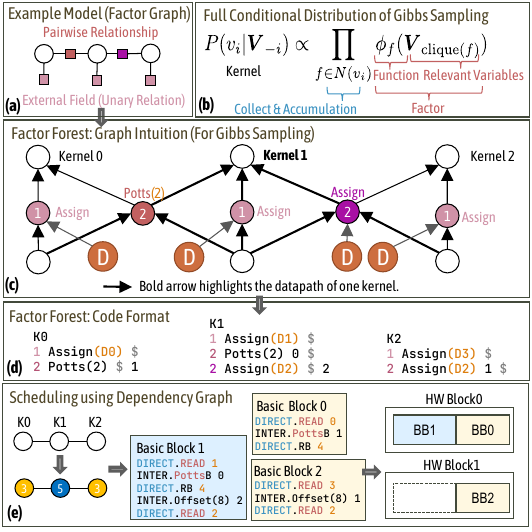}
\caption{
  \ir construction and use: a PGM factor graph is decomposed into per-kernel Factor Forests, encoded in compiler code format, and connected by a Dependency Graph for schedule and hardware-block mapping.
}
\label{fig:ir}
\end{figure}

To enable compiler-driven lowering, we propose \ir, a hardware-agnostic domain-specific intermediate representation for MCMC sampling on probabilistic graphical models. As shown in Fig.~\ref{fig:ir}, \ir is hierarchically composed of two linked abstractions: Factor Forest and Dependency Graph.
The Factor Forest extracts the basic computation units as kernels, while the Dependency Graph connects these kernels to expose dependency-safe scheduling opportunities.
This separation allows the backend to lower local computation into ISA basic blocks and derive schedules and placement decisions from kernel dependencies.


\subsection{Factor Forest}
\paragraph{Semantics \& Syntax}
Based on the widely used factor graphs for representing probabilistic graphical models~\cite{opengm-library,banerjee2019acmc2}, we introduce \emph{Factor Forest} as a computation-oriented representation for conditional-update kernels.
Fig.~\ref{fig:ir}(a) shows an example factor graph with heterogeneous pairwise factors and unary data factors. 
For Gibbs sampling, computing one update kernel requires collecting only the factors connected to the target variable, as reflected by the full conditional subkernel in Fig.~\ref{fig:ir}(b).
The Factor Forest in Fig.~\ref{fig:ir}(c) represents this computation as a collection of computation trees organized as a directed acyclic graph (DAG). 
Fig.~\ref{fig:ir}(d) shows the code format of Factor Forest used by the compiler.

In Factor Forest, each tree corresponds to one update kernel.
Leaf nodes represent relevant current variable states or operand data, while internal nodes represent factor operations on the incoming edges, annotated with functional attributes.
Arguments marked with \texttt{\$} denote the variable being updated and are vectorized over all possible label values.
The root node accumulates incoming vectorized factor contributions to form the probability vector for the corresponding update kernel.
For example, the root node \texttt{K0} is evaluated from a unary data factor \texttt{Assign(D0)} over the candidate labels and a pairwise \texttt{Potts} factor with weight 2 that depends on the candidate label and the current label of variable 1.
Here, \texttt{Assign} reads from data vector \texttt{D0} according to the candidate label, and \texttt{Potts} returns the logical \texttt{NEQ} result over its inputs.

Compared with the original factor graph that is descriptive rather than executable, Factor Forest makes intra-kernel computation patterns explicit.
For sparse probabilistic graphical models, each Gibbs update depends only on the local Markov blanket of the target variable rather than the entire model. Factor Forest exposes this locality by representing each probabilistic distribution as an accumulation of minimal relevant factor contributions.
It also exposes low-level parallelism: the vectorized candidate-label dataflow naturally maps to label-level SIMD execution.

\paragraph{Extensibility} 
Although Factor Forest is inspired by the variable-centric update pattern of Gibbs sampling, it can be extended to other MCMC algorithms that evaluate factorized probability or energy functions.
For example, the exchange step in PT evaluating the global energies of different replicas can be represented as a kernel of Factor Forest whose root accumulates factor contributions across the model. 
The cluster update in BGS can be represented as a kernel whose root node accumulates factor contributions from a block of variables.
The propose-and-acceptance process in the Metropolis-Hastings (MH) algorithm~\cite{calderhead2014general} can be represented as two kernels: one for computing the acceptance probability and another for sampling the accept/reject decision. 

The concrete implementation of factor functions is further optimized during ISA lowering, as discussed in Sec.~\ref{subsec:isa_design}.

\subsection{Dependency Graph}

\paragraph{Semantics}
The Dependency Graph represents update kernels as vertices, each linked to its corresponding Factor Forest tree.
An edge is added between two kernels if the target variable of one kernel is used in the conditional computation of the other.
Such an edge indicates that the two kernels should not be updated concurrently under exact parallel Gibbs sampling.
The graph is undirected because the dependency represents a mutual exclusion constraint for parallel scheduling, rather than a directed execution order.
Avoiding concurrent execution of dependent kernels preserves the validity of the parallelized Gibbs transition.

\paragraph{Usage}
The Dependency Graph exposes both dependency constraints and scheduling opportunities for compiler optimization.
As shown in Fig.~\ref{fig:ir}(e), once Factor Forest kernels are lowered into \codename{} ISA basic blocks, the compiler can derive executable schedules from the Dependency Graph.
For example, graph coloring can identify groups of independent kernels, where kernels with the same color are executed in the same synchronous layer called ``color phase'' and are dispatched to different parallel hardware blocks.

Existing approaches commonly rely on graph coloring to extract parallelism~\cite{banerjee2019acmc2,aadit2022massively,zhao2024aia,zhao2025aia,ko2019accelerating}.
Unlike prior work that treats coloring mainly as an algorithm-level sampling optimization, \ir treats dependency analysis as a compiler optimization problem.
Beyond reproducing coloring-based schedules on the factor graph, the Dependency Graph provides a more expressive abstraction by explicitly encoding kernel-level dependencies and linking to the Factor Forest.
This enables the coexistence of multiple levels of parallelism, as dependency analysis can also be applied to block sampling and parallel tempering.
This also allows auxiliary information, e.g., computational cost, memory access patterns, to be attached to each kernel for backend scheduling and placement optimization.


\section{Back-End: ISA Lowering and Scheduling}
\label{sec:backend}

The backend compiler translates the target-independent \ir into executable control for hardware acceleration.
In the current system, this layer targets the \codename{} ASIC through a custom ISA, pipeline-aware lowering rules, and dependency-aware scheduling metadata.
Because \ir captures coarser-grained semantics before a concrete accelerator mapping, it separates high-level factor representation and scheduling analysis from the AMARA-specific instruction format, memory mapping, and timing constraints.

\subsection{ISA Lowering for Factor Computation}
\label{subsec:isa_design}

%

Within the backend, ISA lowering maps Factor Forest operations to compact instructions that drive factor computation on the \codename{} accelerator, as shown by the example in Fig.~\ref{fig:ir}(d)--(e).

\begin{figure}
\centering
\includegraphics[width=\linewidth]{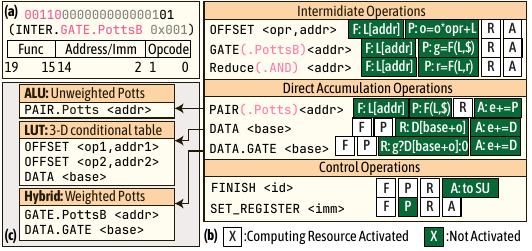}
\caption{\codename{} ISA and compiler lowering.
(a) The RISC-like fixed-format instruction of \codename{}.
(b) Representative instructions activate selected subsets of the pipeline resources.
(c) The compiler lowers different example factors to the corresponding instruction templates.}
\label{fig:ISA}
\end{figure}

\subsubsection{\codename{} ISA Format}

The ISA uses a fixed-length, RISC-like format, but adopts coarse-grained operations tailored to factor-based computation in probabilistic sampling.
Factor operation, memory access, and accumulation behavior are encoded through three fields: \verb|Func|, \verb|Addr|, and \verb|Opcode|, as illustrated in Fig.~\ref{fig:ISA}(a).

Instructions are categorized into three groups:
\texttt{INTER} instructions compute intermediate results for factor functions that require multi-step evaluation, such as high-arity or offset-based factors.
\texttt{DIRECT} instructions directly accumulate factor contributions into the probabilistic distribution value of the current update kernel.
\texttt{CTRL} instructions control sampling-related operations and machine states.

Designed for streamlined dataflow, the ISA omits branching instructions.
Source and destination operands are implicitly determined by the \verb|Func| and \verb|Opcode| fields, removing the need for explicit register specification.
Furthermore, memory load/store operands are encoded in the same instruction as factor computation: the \verb|Addr| field selects the required label-memory or data-cache entry.
Consequently, each instruction specifies the factor operation, input source, and destination behavior using only a function code and one address/immediate field, reducing instruction count and simplifying backend code generation.

\subsubsection{Co-designed Pipelined ISA}
\label{subsubsec:isa_pipe}
\codename{} co-designs its pipelined ISA with factor-function lowering. The ISA exposes four stage execution resources: label-state fetch (F), arithmetic/logic processing (P), factor-data read (R), and accumulation (A).
Each opcode activates only the resource subset required by its operation; for example, ALU-only instructions use F/P/A, whereas table-defined factors additionally use R.
Fig.~\ref{fig:ISA}(b) summarizes the representative operations and their execution semantics.

This restricted execution model simplifies both compilation and hardware control.
The backend selects a fixed instruction template according to the factor-function type instead of performing general instruction scheduling over arbitrary register dataflow. Most factor contributions are independent; only multi-step factor functions communicate through the dedicated intermediate registers defined by the ISA.
Consequently, the EU executes template-conformant streams without dynamic instruction-dependence tracking, forwarding, or instruction-level interlocks. 



\subsubsection{ALU-LUT Hybrid Lowering}
\label{subsec:isa_lowering}

Factor functions in Factor Forest are lowered to compact ISA instruction sequences through an ALU--LUT hybrid strategy.
ALU operations are used for simple symbolic computation, while data-cache lookup is used for parameterized or table-defined factor values.
This hybrid design covers both arithmetic and table-defined factors without requiring a large universal lookup table for every factor function.
Fig.~\ref{fig:ISA}(c) shows representative templates for the ALU-only, LUT-only, and hybrid cases.

An unweighted Potts factor uses the fused \texttt{PAIR.Potts} template.
A table-defined conditional factor uses repeated \texttt{OFFSET} operations to construct a table address, followed by \texttt{DATA} to read and accumulate the selected entry.
A weighted Potts factor first uses \texttt{GATE.PottsB} to form a label-dependent predicate, then uses \texttt{DATA.GATE} to read and accumulate the weight only when the predicate holds.
The hybrid template therefore combines symbolic evaluation with parameter lookup without fully materializing the weighted function as a LUT.

\subsubsection{Data Precision Tuning}
\label{subsubsec:prec_tune}
The backend converts factor values to the log domain and selects a fixed-point format for each compiled workload.
It analyzes lower and upper bounds of the factor contributions in every update kernel tree to estimate the possible span of the log-likelihood value, from which it allocates the required signed integer bits.
The remaining fractional width is a workload-dependent accuracy--storage trade-off; integer-valued optimization workloads can use zero fractional bits.

The compiler may also center factor values to reduce the required dynamic range. Such an offset must be common to all candidate labels of the affected update so that it preserves the relative log-likelihoods consumed by the sampler. The resulting format is encoded in compiled factor data and used consistently by the EU accumulator and SU.

\subsection{Dependency-Aware Scheduling}
\label{subsec:compiler_opt}

Given a Dependency Graph, per-kernel cost estimates, and a target number of processing blocks, the backend generates a static placement and an ordered phase schedule. The schedule preserves dependency constraints while the placement and phase assignments seek to reduce cross-block communication and phase-level under-utilization.

This optimization is required by the inherent tension between dependency constraints and parallel execution.
On one hand, proposal kernels in component-wise MCMC must respect the sequential transition semantics imposed by variable dependencies to ensure ergodicity. 
On the other hand, modern hardware requires sufficient parallelism and locality to achieve high utilization.
Naively executing proposals either sacrifices correctness (by violating dependencies) or performance (by under-utilizing parallel resources and incurring excessive memory communication).

\subsubsection{Dependency-Safe Phase Scheduling}

A coloring $c:V\rightarrow\{1,\ldots,C\}$ satisfies $c(u)\neq c(v)$ for every $(u,v)\in E$; therefore, each color class is an independent set of update kernels.
During a sweep, \codename{} executes the kernels of one color phase in parallel and commits their sampled labels before advancing to the next phase. No two mutually dependent kernels are active in the same phase.

\subsubsection{Partition-Aware Balance Optimization}

We assign each update kernel $v$ to one of $B$ processing blocks through a placement $p:V\rightarrow\{1,\ldots,B\}$. 
The placement guides cross-block communication and the label-memory mapping described in Sec.~\ref{subsec:mem}.
However, balanced total cost does not imply balanced cost within every color phase. Because the most heavily loaded block determines phase completion time, other blocks can remain idle at the phase barrier.

To address this, we adopt a coupled two-stage optimization method:
The first stage applies weighted METIS partitioning, using estimated kernel costs to balance total block workload while discouraging dependency edges that cross blocks.
With the placement fixed, the second stage applies a partition-aware variant of the DSATUR heuristic coloring. For each color $c$, let $\ell_{c,b}$ denote the workload assigned to partition $b$ in color $c$. Among legal color assignments, the heuristic prefers assignments that reduce $\mathrm{SSE}(c)$.
\begin{equation}
\mathrm{SSE}(c)=\sum_b \ell_{c,b}^2-\frac{\left(\sum_b \ell_{c,b}\right)^2}{B},
\end{equation}

The objective of graph coloring is therefore not to minimize the chromatic number, but to construct legal phases with sufficient parallelism and lower block-level imbalance.
The two-stage design avoids introducing complex constraints into the partitioning step and provides an effective heuristic for phase-aware scheduling.


\section{Compiler-Driven \codename Accelerator}
\label{sec:arc}

\begin{figure}[t!]
\centering
\includegraphics[width=\linewidth]{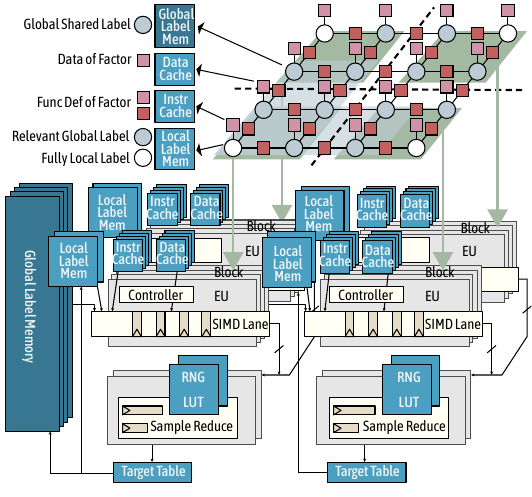}
\caption{Block diagram of the overall architecture of \codename, onto which PGMs are straightforwardly deployed.
}
\label{fig:Architecture}
\end{figure}

Because \ir exposes both the local factorized computation and the inter-kernel dependency structure of a workload, a hardware architecture designed around these abstractions can execute compiler-generated schedules without runtime overhead such as dynamic dependency resolution or general-purpose instruction decode.
This observation motivates the \codename{} domain-specific accelerator, whose datapath, memory hierarchy, and control logic are tailored to the instruction templates and phase-based schedules produced by the backend (Sec.~\ref{subsec:compiler_opt}).
Concretely, the compiler maps each logical partition of the Dependency Graph to one physical processing block; each block then consumes its phase-ordered instruction stream and label working set (Fig.~\ref{fig:Architecture}).

Within a block, a controller dispatches instructions and data to \textit{energy unit}s (EUs), which evaluate candidate-label probabilities; \textit{sampling unit}s (SUs) then sample new labels and write them to the block's banked local label memory. Labels required by multiple blocks are selectively made visible through the shared global label memory.
The following subsections detail the EU datapath, SU datapath, and phase-aware label-access protocol, respectively.

\subsection{EU Design}
\label{subsec:EU}

\begin{figure}[t]
\centering
\includegraphics[width=\linewidth]{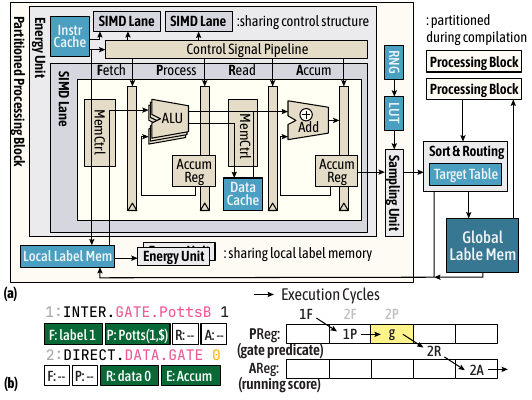}
\caption{(a) Hierarchical design of the processing block, containing several EUs supporting SIMD execution. (b) The pipeline timing illustration for the hybrid instruction template example.}
\label{fig:EU}
\end{figure}

As shown in Fig.~\ref{fig:EU}(a), each EU evaluates the compiler-generated factor instructions for one update kernel and produces an unnormalized log-likelihood score for every candidate label.
SIMD lanes share the instruction stream while holding distinct candidate labels, thereby evaluating the candidate-label score vector in parallel.
Independent update kernels and Markov chains can be mapped across EUs and processing blocks according to the compiler-generated schedule.



\paragraph{Accumulation-Centric Computation} 

The root of a Factor Forest aggregates factor contributions for an update kernel.
The EU therefore uses an accumulation-centric datapath instead of a general register file, matching the reduction structure of the \ir while avoiding general register movement.
In each computing lane, two types of accumulative registers are implemented.

The intermediate accumulation registers are introduced to hold temporary states for high-order factor functions that cannot be computed through basic operators in the ALU in one cycle and are the target of the \texttt{INTER} instructions.
The energy accumulation registers hold the running score for each candidate-label lane as the probability result, and are the target of the \texttt{DIRECT} instructions. The accumulation results in the EUs directly correspond to the root node in the Factor Forest.

\paragraph{Pipeline Implementation}

Each SIMD lane physically realizes the F/P/R/A execution model introduced by the co-designed ISA in Sec.~\ref{subsubsec:isa_pipe}, as shown in Fig.~\ref{fig:EU}(a).


F accesses label memory, P evaluates symbolic operations or generates LUT offsets, R accesses factor data when required, and A adds a completed contribution to the lane's running score.
F stage accesses label memory.
P stage performs arithmetic and logic operations for ALU-based factor evaluation, and generates offsets for LUT-based computation.
R stage accesses the data memory when required.
A stage accumulates the computed factor value into the log-likelihood (energy) and forwards the result to the sampling unit.

For a hybrid factor, the temporary state produced by an \texttt{INTER} instruction is retained locally and consumed by the following \texttt{DIRECT} instruction. Thus, the stage order organizes the intra-factor flow of a template; it does not impose operand dependencies among distinct factor contributions. Fig.~\ref{fig:EU}(b) illustrates this case.
Because the compiler emits only these fixed factor templates, the EU needs no dynamic instruction-dependence tracking.


\paragraph{Numerical Co-design}
The compiler applies the log-domain conversion, offset, and the fixed-point encoding described in Sec.~\ref{subsubsec:prec_tune} before factor data are placed in the memories.
The EU then accumulates these encoded factor contributions using fixed-point addition and produces an unnormalized log-likelihood score vector.
This format is also the EU--SU interface: the SU consumes the score vector directly for Gumbel-Max sampling, without probability normalization or numerical-format conversion between the units.

\subsection{SU Design}
\label{subsec:SU}

\begin{figure}[t]
\centering
\includegraphics[width=\linewidth]{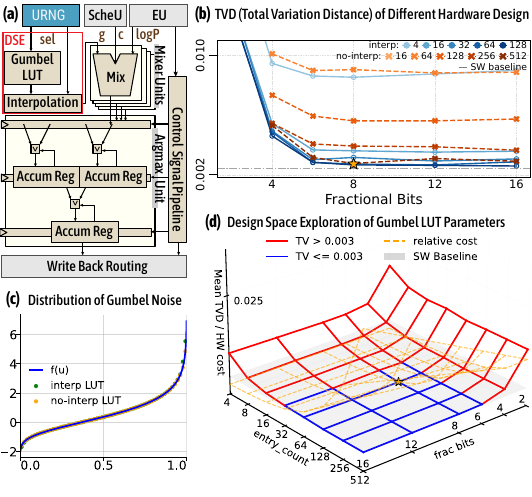}
\caption{Components of Sampling Unit and relevant data support.}
\label{fig:method-sampler}
\end{figure}

The SU transforms the EU-generated log-likelihood values into discrete samples drawn from the corresponding softmax distribution.

A traditional softmax sampler requires exponential evaluation, overflow prevention, normalization, and cumulative distribution function (CDF) construction, which introduce serialization and limit pipelining.
\codename{} instead applies the Gumbel-Max reparameterization~\cite{zhao2025mc2a}, converting sampling into adding random perturbations to the log-likelihood probability scores and maximum selection over the scores.
The Gumbel-Max sampler supports streaming sampling from a probability distribution of arbitrary length through temporal accumulation of the argmax operation.

\paragraph{Sampler Implementation}

\begin{algorithm}[ht]
\caption{Tempered Argmax Sampler}\label{al:1}
\KwData{logarithmic probability $\boldsymbol{L}$ with $l$ components, temperature parameter $c$}
\KwResult{sample $s\sim p,\,p_i=\operatorname{softmax}(\boldsymbol L/T)_i$ }
\Begin{
\ForAll{$i \in \{1,\dots,l\}$ \textbf{in \textit{parallel}}}
{
generate $u_i\leftarrow \mathrm{UniformRandom}(0,1)$\;
lookup noise $g_i\leftarrow -\log(-\log(u_i))$\;
mix score $m_i \leftarrow g_i\cdot c+L_i\cdot(1-c)$\;
}
\Return{$s\leftarrow \arg\max \{m_1,\dots,m_l\}$}
}
\end{algorithm}

The Gumbel-Max SU uses a LUT-based Gumbel noise generator with interpolation (Fig.~\ref{fig:method-sampler}(a) and Algorithm~\ref{al:1}).
The sampler datapath consists of the mixer and the argmax unit.
After receiving the Gumbel perturbations from the interpolated LUT, an argmax accumulator maintains the maximum mixed score and its label index as candidate probability scores stream through the SU.

The mixer combines Gumbel noise $g_i$ with log-likelihood values $L_i$ under the control of a coarse-grained temperature parameter $c\approx\frac{T}{T+1}$, implemented with low bit-width, and the mixing operation is mapped to a shallow adder tree with simple fixed-point arithmetic.

\paragraph{Numerical Performance}
Gumbel Noise is numerically safe. The represented range of \(g_i\) spans eight bits using a 64-entry LUT, see Fig.~\ref{fig:method-sampler}(c).

The numerical robustness of the proposed sampler is validated through both distribution-level and system-level evaluations (Fig.~\ref{fig:method-sampler}(b)(d)).
Fig.~\ref{fig:method-sampler}(b) evaluates the Total Variation Distance (TVD) between the hardware sampler and the ideal softmax distribution under different quantization configurations. Increasing fractional bit-width significantly reduces TVD, with diminishing returns beyond 8 bits. Moreover, interpolation consistently improves accuracy versus non-interpolated LUT designs, especially under low-precision settings. 

Fig.~\ref{fig:method-sampler}(d) presents a joint exploration over LUT entry count and fractional precision. The results show a clear trade-off between hardware cost and statistical accuracy. Configurations with moderate LUT size (e.g., $64\sim 128$ entries) and $8\sim 12$ fractional bits achieve low TVD while keeping hardware overhead close to the software baseline. The highlighted operating point demonstrates an efficient balance, delivering near-optimal accuracy with limited cost.

\subsection{Memory Design}
\label{subsec:mem}


\codename{} employs a hierarchical and distributed memory system to support the parallel execution of EUs and SUs while avoiding a centralized bandwidth bottleneck.
The memory system comprises instruction memory for compiled control, data memory for factor parameters and tables, and label memory for the mutable sampler state.
\label{position:share-concern-2-1}
\re{In this subsection, all memory designs are realized as on-chip banked SRAM.
}

\paragraph{Distributed Memory System}

The data memory and the instruction memory are statically scheduled during execution. During a sweep of component-wise MCMC kernels, they are sequentially scanned and are therefore easy to prefetch and buffer as a scratchpad.
Instruction memory is distributed across EUs, and data memory is distributed across SIMD lanes in the EUs.
In contrast, label states are updated continuously during MCMC sampling and therefore require a dedicated memory organization. \codename{} adopts a hierarchical label memory design composed of block-level local label memories and a shared global label memory.

This block-level sharing strikes a balance between bandwidth and communication cost.
On one hand, it provides high-bandwidth access for EUs within the block through a banked SRAM structure with address decoders and arbiters, enabling concurrent accesses from multiple EUs.
On the other hand, it avoids excessive replication of label states across all EUs, which would otherwise increase storage overhead and complicate synchronization.

The local label memory is banked and accessed through address decoders and arbiters, allowing multiple SIMD EUs in the same block to fetch label states concurrently.
Both local and global label memories are organized into multiple banks to support concurrent read and write operations. Compared with prior designs that rely on full crossbar interconnects to support all-to-all access among distributed register banks~\cite{zhao2025mc2a,shah2022dpu}, \codename{} reduces communication complexity from $O(N^2)$ to $O(N)$.
During execution, each proposal update is first written to the local label memory of its corresponding block. The update is selectively published to the global label memory only when required by other blocks.

\paragraph{Placement and Access Optimization}

The hierarchical label memory design provides high access bandwidth through distributed local memories and maintains cross-block label visibility through a shared global memory.
However, this hardware organization alone does not specify when globally visible labels should be fetched, reused, or synchronized during execution.
Without careful management, accesses to globally visible labels can still become a memory bottleneck when parallel EU blocks scale up.
Private placement that avoids replica coherence issues repeatedly reads nonlocal labels from global memory, causing a major bottleneck in dense graphs. \codename{} instead replicates the working set required by each block, reducing these reads while requiring a phase-aware freshness mechanism.

\codename{} leverages the phase-based execution induced by graph coloring to eliminate the need for fine-grained coherence.
Since each variable is updated in a specific phase and remains unchanged until the next sweep, the validity of a label can be determined using phase information rather than explicit coherence checks.
Specifically, each label entry in local memory is augmented with an extra field that marks whether it has been accessed after its last update phase. When a variable is first accessed after its update phase, its value is fetched from global memory and stored locally. Subsequent accesses within the same sweep are served entirely from local memory.
This design establishes a read-once invariant: each variable is fetched from global memory at most once per sweep, reducing global memory traffic and improving scalability for high-density models.



\section{Evaluation}
\label{sec:evaluation}

\subsection{Experimental Setup}

\subsubsection*{Implementation}
\label{subsubsec:impl}
We implement \codename{} in SystemVerilog and synthesize it using Synopsys Design Compiler with a standard-cell library targeting TSMC 5nm FinFET Plus (5FFP) process technology. Sampling throughput is measured through cycle-accurate RTL simulation.
The synthesized implementation of \codename{} has 128 EUs and 32 SUs. Each EU has two SIMD lanes generating 16-bit fixed-point energy values, two 1024$\times$16-bit Data Caches, and a 1024$\times$32-bit Instruction Cache. 4 EUs form a block sharing the same 8192$\times$8-bit local label memory and the same 8192$\times$16-bit target table. The global label memory is 8192$\times$8-bit.
The URNG is generated from a PIM-based design that efficiently outputs RRNG numbers~\cite {fu2024pcim}. Each Gumbel noise LUT has 64 entries and 12-bit word width. The width of the temperature parameter is 6 bits.
\label{position:design-capacity}
\re{
This design covers all evaluated benchmarks and can store up to 262,144 random variables across its local label memories. For instance-structured models such as COP, the tractable problem scale is about $10^4$ variables~\cite{yue2024scalable,li2026ising}. 
}

\label{position:design-parameter}
\re{The number of SIMD lanes is set to 2 to maximize the hardware utilization, considering some of the workloads modeled as QUBO problems have a small number of labels to expose the label-level parallelism. The sampling throughput can be further improved by increasing the number of SIMD lanes when processing multi-state models or multi-chain workloads. Considering the computing component of the AMARA architecture is not the main source of the hardware overhead, the adaptation of the SIMD lane number is also favorable.}
\re{The EUs number per one block is a practical compromise between performance and hardware overhead. 
We evaluate this trade-off by varying the block size from 1 to 16 while keeping the total EU count and the label memory capacity constant. 4 EUs per block provide $1.39\times$ geometric-mean throughput relative to one EU per block. Increasing the block size from 4 to 16 yields a further 10.6\% throughput improvement, while the aggregate local label memory router cell count increases by $4.38\times$.
}


\subsubsection*{Benchmarks}


We evaluate \codename{} on 11 representative probabilistic graphical models.
The \textit{Pegasus} and \textit{Zephyr} graphs, developed for D-Wave quantum annealers, can serve as fixed topologies for sparse deep Boltzmann models~\cite{niazi2024training,dattani2019pegasus,boothby2021zephyr}.
\textit{Alarm}, \textit{Water}, and \textit{Andes} are medium-scale Bayesian networks.
\textit{G6}, \textit{G22}, \textit{G49} from Gset serve as Max-Cut QUBO instances.
\textit{Queen13} is a graph-coloring instance from the COLOR benchmark, formulated here as an integer-state optimization problem~\cite{trick2002color}.
\textit{uf100} and \textit{uf250} are satisfiable random 3-SAT instances from SATLIB, which introduce high-order relationships~\cite{hoos2000satlib}.

\subsubsection*{Baselines}
\label{position:baselines}
The software experiments are conducted on a platform equipped with an Intel Xeon Platinum 8470Q CPU @2.10~GHz and an NVIDIA GeForce RTX 5090 GPU @2.01~GHz.
\re{Three different software backends are used for comparison: Numba JIT backend implementation of Gibbs Sampling on CPU and GPU (Numba); software library thrml using the JAX backend (JAX); lowered MCMC-IR to GPU using MLIR (MCMC-IR). 
The software implementation enables kernel-level parallelism using the static DSATUR coloring and chain-level parallelism by replicating states and running multiple independent Markov chains in parallel.}

\re{We also evaluate two programmable sampling architectures: AIA~\cite{zhao2025aia} is a 16-core RISC-V SoC with custom Knuth--Yao sampling instructions and inter-core register sharing. \codename{}-cv replaces \codename{}'s EU/SU execution with CV32E40P cores~\cite{openhw2026cv32e40p} while retaining its memory organization and compiled partitions and phases. MCMC-IR targeting and scheduling methods are enabled on them.}




\subsection{Hardware Overhead}

\begin{table}[t]
    \centering
    \caption{Power \& Area Breakdown of \codename{} with 32 Blocks}
    \begin{tabularx}{\columnwidth}{l | X | X | X}
    \hline
      \textbf{Component} & \textbf{Quantity} & \textbf{Power} & \textbf{Area} \\
                         &                    & \textbf{/mW}   & \textbf{/mm$^2$} \\
      \hline
      \multicolumn{4}{c}{\textit{Computing Components}} \\
      \hline
      EU & 128 & 42.858 & 0.055619 \\
      SU & 32 & 9.824 & 0.010860 \\
      \hline
      \multicolumn{4}{c}{\textit{Memory Components}} \\
      \hline
      URNG Module & 0.5\,KB & 1.801 & 0.011264 \\
      D\& I Memory & 1024\,KB & 215.115 & 0.057430 \\
      Label Memory & 1040\,KB & 202.112 & 0.053122 \\
      \hline
      \textbf{Total} & 2064.5\,KB & 471.710 & 0.188295 \\
      \hline
    \end{tabularx}
    \label{tab:pa_break}
\end{table}

Table~\ref{tab:pa_break} presents the specific architecture configuration, power consumption, and area of a full \codename{} design (introduced in Sec.~\ref{subsubsec:impl}) synthesized in TSMC 5FFP process technology.
The power and area of the computing components are estimated from the post-synthesis netlist, while the memory components are estimated using an SRAM-based analytical model parameterized by memory capacity and data width.

The on-chip memory components, including the global and label memory, data cache, and instruction cache, account for a significant portion of the total power and area.
In most listed benchmark cases, the memory is underutilized because the model size is much smaller than the set maximum capacity.

\subsubsection*{\re{Modeled off-chip bandwidth and prefetch feasibility}}
\label{subsubsec:dram_modeling}

\begin{figure}[t]
    \centering
    \includegraphics[width=\linewidth]{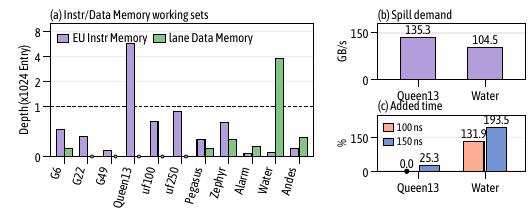}
    \caption{
      \re{
        (a) Usage of the Instr/Data Memory compared to the 1024-entry on-chip capacity. (b) The external support demand of overflowing cases. (c) Modeled steady-state time increase.
      }
    }
    \label{fig:external-memory}
\end{figure}
\label{position:share-concern-2-2}
\re{
The reported performance of \codename{} results assume that the model fits within the on-chip SRAM capacity (Table~\ref{tab:pa_break}).
The data and instruction working sets may exceed the on-chip SRAM capacity, requiring off-chip DRAM access; see Fig.~\ref{fig:external-memory}(a).
At the current size setting, the modeled off-chip bandwidth requirement is up to 573.44 GB/s assuming full hardware utilization; on the measured workload, the actual bandwidth requirement is lower, see Fig.~\ref{fig:external-memory}(b).}

\re{The effect of this off-chip access is revealed by simulation tests including DRAM. For a modeled two-stack HBM2E interface, the sequential prefetching can solve the capacity issue for \textit{Queen13}; \textit{Water}'s large data table limits prefetch effectiveness (Fig.~\ref{fig:external-memory}(c)), but this can be mitigated by other cache techniques such as LRU.
Label memory is within the on-chip capacity for all evaluated benchmarks.
}

\subsection{Performance}

\subsubsection{Compilation Time}
\begin{figure}
    \centering
    \setlength{\abovecaptionskip}{3pt}
    \includegraphics[width=\linewidth]{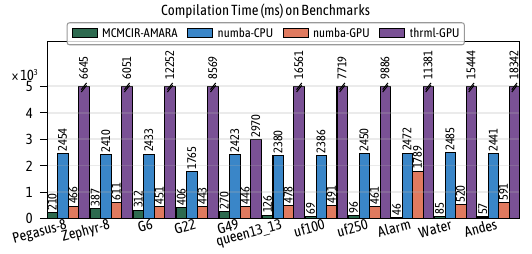}
    \caption{Compilation Time of \codename{} and software implementations}
    \label{fig:eval-compT}
\end{figure}

Fig.~\ref{fig:eval-compT} shows the comparison of compilation time for employing the sampling workloads on different platforms. The software backend implementation parses \ir and enables chain-level parallelism by replicating states and kernel-level parallelism from the given result of original DSATUR coloring. The \ir to \codename{} compilation time includes metis partitioning and partitioning-aware coloring.
The programming library thrml with JAX GPU backend uses substantially more compilation time. The \ir-to-\codename{} path is shorter because it directly maps the \ir to the \codename{} ISA, reflecting the co-design of the IR and the architecture.

\subsubsection{Sampling Throughput}
\begin{figure}
    \centering
    \setlength{\abovecaptionskip}{3pt}
    \includegraphics[width=\linewidth]{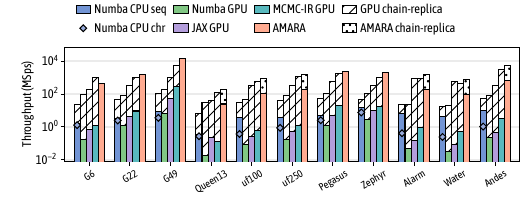}
    \caption{Measured throughput of various PGM workloads for hardware accelerators and different hardware backends on programming libraries.}
    \label{fig:eval-throughput}
\end{figure}

Throughput is measured by the number of transition samples generated per second, using the unit MSps (million samples per second). CPU and GPU implementations without chain-level parallelism give poor performance because of the limited kernel-level parallelism. GPU behaves even worse because the control overhead for chromatic sampling outstrips the benefit from parallel computing.

The shaded bars in Fig.~\ref{fig:eval-throughput} represent the sampling throughput enabling independent chain-level parallelism, i.e., replicating the Markov chain.
This chain-level parallelism brings an increase in the throughput, but the throughput gain is lower than the increase in the degree of parallelism.

The throughput of \codename{} is measured through cycle-accurate simulation on the hardware implementation described in Sec.~\ref{subsubsec:impl}. For \textit{G6}, \textit{G22}, \textit{G49}, \textit{Pegasus}, and \textit{Zephyr}, only kernel-level parallelism is enabled. For the remaining benchmarks, the scale of the model is relatively small, so the kernel-level parallelism is limited to 16 EUs. The remaining parallelism is realized by replicating the Markov chain and is marked with shading. On average, the implemented version of \codename{} achieves $53.1\times$ and $6.1\times$ speedup over the maximum multi-chain CPU and GPU throughput, respectively. \re{Across 11 workloads, the MCMC-IR GPU backend outperforms JAX GPU on nine and achieves a $2.86\times$ geometric-mean throughput gain, while AMARA is $1.66\times$ faster than MCMC-IR GPU on average.}
\label{position:mcmc-ir-gpu}

\subsubsection{\re{Sampling Quality}}
\label{subsubsec:sampling_quality}

\begin{figure}[t]
    \centering
    \includegraphics[width=\linewidth]{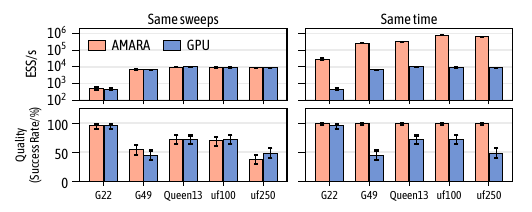}
    \caption{\re{Sampling quality comparison of \codename{} and GPU software implementation.}}
    \label{fig:eval-quality}
\end{figure}

\re{
Sec.~\ref{subsec:SU} describes the numerical accuracy of the Gumbel-Max sampler with quantization and interpolation. Here we evaluate the overall sampling quality of \codename{} in terms of effective sample size per second (ESS/s) and optimization quality for combinatorial optimization problems, as shown in Fig.~\ref{fig:eval-quality}.
With the same sweep count, \codename{} and the software baseline behave essentially the same.
AMARA achieves about 60$\times$ improvement in ESS/s across four binary workloads and increases optimization success rates by up to 55.5 percentage points under matched runtime budgets.
}

\subsection{Optimization Validity}
\label{subsec:ablation_profile}

\subsubsection{\re{Programmable Execution Path}}
\label{subsubsec:exec_study}

\begin{figure}[t]
    \centering
    \includegraphics[width=\linewidth]{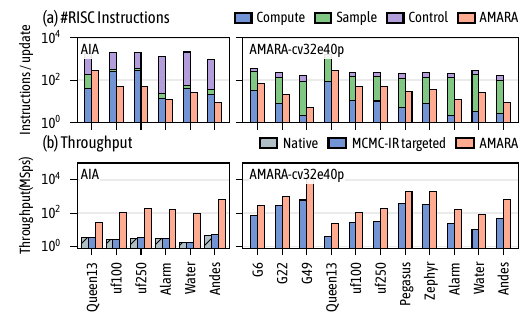}
    \caption{\re{(a) Instruction count and (b) throughput comparison of AMARA and two RISC-V-based backends. Total bar height of (a) uses a logarithmic scale, while colored segment heights represent linear instruction shares.}}
    \label{fig:risc-comparison}
\end{figure}

\re{
  Fig.~\ref{fig:risc-comparison} compares instruction counts and throughput against two programmable execution paths. With matched computing cores, \codename{} achieves
$3.27$--$23.43\times$ higher throughput than AMARA-cv and $7.40$--$132.50\times$ higher than AIA. 
The instruction profiles illustrate the reduced overhead enabled by compact computation instructions and hardware sampling.
}

\subsubsection{Memory Conflict \& Locality}

\begin{figure}
    \centering
    \setlength{\abovecaptionskip}{3pt}
    \includegraphics[width=\linewidth]{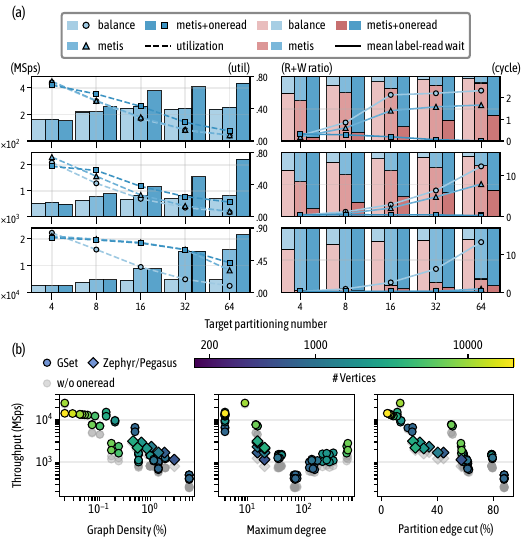}
    \caption{(a) Study on scheduling global label memory. From top to bottom are the results on G6, G22, and G49. \re{(b) The performance of AMARA under different graph statistics.} }
    \label{fig:eval-sche}
\end{figure}

Fig.~\ref{fig:eval-sche}(a) analyzes the memory-bound problem when increasing the kernel-level parallelism. The three graphs taken from GSet have different densities and scales. \textit{G6} is a small and dense graph, \textit{G22} is a medium-scale graph with medium density, and \textit{G49} is a large and sparse graph.
The simulation results of workload throughput measured by MSps and profiling information, including utilization of the EU, ratio of local and global memory access, and average waiting time for label access, are shown for three methods: \textit{balance} enforces load-balanced partitioning without considering cut costs; \textit{metis} uses composite scheduling with METIS partitioning but without the read-once technique. \textit{metis+oneread} is the full optimization method.

\begin{table*}[t!]
\centering
\caption{Qualitative comparison with representative probabilistic-computing accelerators
}
\scriptsize
\setlength{\tabcolsep}{1.5pt}
\renewcommand{\arraystretch}{0.95}
\begin{tabularx}{\textwidth}{@{}>{\raggedright\arraybackslash}p{0.115\textwidth}*{7}{|>{\centering\arraybackslash}X}@{}}
\hline
 & AcMC$^2$\cite{banerjee2019acmc2} & PROCA\cite{fu2025proca} & ESTroM\cite{fan2026estrom} & BigLittleMCA\cite{kjellqvist2025biglittle} & AIA\cite{zhao2025aia} & CoopMC\cite{chai2022coopmc} & \codename (This Work) \\
\hline
Platform & FPGA/ASIC & ASIC & ASIC & ASIC & ASIC & ASIC & ASIC \\
Technology & 28nm & 28nm & 28nm & 16nm & 16nm & 12nm & 5nm \\
Frequency (MHz) & 400 & 200 & 500 & 3000 & 300 & 500 & 1000 \\
Peak Throughput$^\circ$
& $2.3\times10^{9}$ & $1.3\times10^{7}$ & --\textsuperscript{\textdagger} & $10^{10}$ 
& $1.2\times10^{9}$ & $3.1\times10^{8}$ & $3\times10^{10}$ \\
\re{Area Efficiency$^\bullet$}
& \re{--} & \re{$2.28\times10^3$} & \re{--} & \re{$3.81\times10^3$} & \re{$3.07\times10^3$} & \re{$8.98\times10^4$\textsuperscript{$\ddagger$}} & \re{$1.1\times10^5$} \\
\hline
Supported Model
& PPL/BLOG PGMs
& Causal PGMs
& Sparse TPMs (PCs)
& Grid MRFs 
& discrete PGMs 
& discrete PGMs 
& discrete PGMs 
\\
Compiler
& PPL HLS
& Custom RISC-V
& Sparsity-aware PC compiler
& Fixed-function
& Custom RISC-V
& Fixed-function
& \ir to RISC ISA \\
Parallel Technique
& Dataflow scheduling
& SIMD, branch prediction
& Element-flow execution
& Fixed spatial tiling
& Fixed Multi-core mesh
& Fixed-length TreeSampler
& Multi-level, compiler-driven 
\\
\hline
\end{tabularx}
\footnotesize\textsuperscript{\textdagger}ESTroM is a DAG accelerator that does not produce probabilistic samples. \textsuperscript{$\ddagger$} Only the core logic area is reported. $^\circ$MSps. $^\bullet$MSps/mm$^2$, normalized to 5nm.
\label{tab:comparison}
\end{table*}

With the increase in the number of partitions for kernel-level parallelism and given the fixed bandwidth of global label memory, the throughput of the global label memory always becomes the bottleneck for \textit{balance} method, so the throughput does not scale with the number of parallel units. METIS partitioning and the read-once memory management technique reduce the requirement for reading global label memory, so the throughput scales better with the number of parallel units. The read-once memory management itself brings some control overhead, so the throughput is slightly lower when data access from global memory is not the bottleneck.

Across the evaluated Gset instances, \texttt{metis} and \texttt{metis-oneread} achieve average speedups of $1.16\times$ and $2.36\times$ with a 32-way partitioning target, respectively. The read-once management is more effective for the dense graph (\textit{G6}) than the sparse graph (\textit{G49}) because the dense graph has a higher chance of reusing the label value update from another processing element. \re{This trend is also shown in Fig.~\ref{fig:eval-sche}(b): the cases with higher density and partition edge cut have a higher gain from the memory management technique.}

\begin{figure}[H]
    \centering
    \setlength{\abovecaptionskip}{3pt}
    \includegraphics[width=\linewidth]{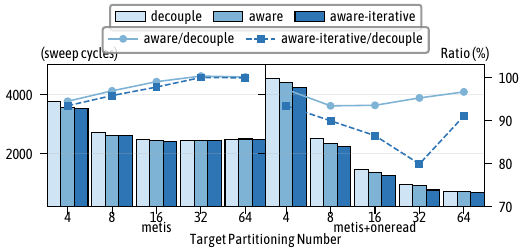}
    \caption{Study on composite scheduling.}
\label{fig:eval-balance}
\end{figure}

\subsubsection{\re{Sensitivity to Graph Characteristics}}
\label{subsubsec:graph_characteristics}

\re{
  Fig.~\ref{fig:eval-sche}(b) analyzes the topology-dependent scalability of \codename{}. The test is conducted on the full set of Gset instances and the \textit{Zephyr} and \textit{Pegasus} graphs with different scales.
  Increasing the vertex count by $4.53\times$ on \textit{Pegasus} and $5.83\times$ on \textit{Zephyr} improves \codename{}'s throughput by $1.87\times$ and $1.91\times$, respectively, showing no throughput degradation with increasing graph size over the evaluated range. At a fixed size of 2,000 vertices, the random graphs have approximately $5\times$ higher density than the four-regular graphs and 84.8\% lower median throughput. The hub-heavy graphs exhibit a smaller throughput reduction of 27.1\%, despite a $6.95\times$ increase in maximum degree and substantially greater degree skew.
}

\subsubsection{Balance}

Fig.~\ref{fig:eval-balance} takes \textit{G22} as an example to analyze the effect and the boundary of the partitioning-aware coloring method. Three methods are compared: \textit{decouple} does graph coloring and partitioning independently; \textit{aware} uses the SSE objective to adjust the original DSATUR coloring result; \textit{aware-iterative} iteratively searches for a lower SSE coloring result until the improvement is smaller than a threshold. Both results with and without the read-once memory management technique are shown.

When the partitioning number is small, the chance of a random bubble by decoupled coloring is small, so the improvement of aware coloring is not significant, as in the case of \textit{metis+oneread} + \textit{aware-iterative}, from 4 to 32 partitions.
When the throughput is bottlenecked by global memory access, the workload imbalance will be smoothed out by the waiting time for the memory access, so the improvement of partitioning-aware coloring is not significant, as in the case of \textit{metis} and the 64-partition example of \textit{metis+oneread}.
Partition-aware iterative coloring improves throughput by 27.7\% on Gset instances at 16 partitions.






\subsection{Comparison with Related Work}

Table~\ref{tab:comparison} compares \codename{} and state-of-the-art probabilistic computing accelerators.
Compared with AcMC$^2$~\cite{banerjee2019acmc2}, which compiles a supported subset of PPL/BLOG models into model-specialized MCMC hardware, \codename{} targets arbitrary-topology discrete PGMs with an explicit IR, programmable ISA, and architecture-aware mapping flow.
PROCA~\cite{fu2025proca} and ESTroM~\cite{fan2026estrom} target causal inference and sparse tractable probabilistic circuits, respectively.
PROCA cross-compiles PyMC posterior-PDF functions for causal-inference execution, and ESTroM maps a
sparse probabilistic-circuit DAG into node- and edge-flow computation. These designs efficiently exploit the structure of their target model families, whereas \codename{} targets discrete PGMs with arbitrary, instance-defined topology. 
Compared with BigLittleMCA~\cite{kjellqvist2025biglittle}, AIA~\cite{zhao2024aia} and CoopMC~\cite{chai2022coopmc}, \codename{} provides a compiler-driven optimization flow.
They primarily specialize the sampling execution substrate: BigLittleMCA exploits a fixed mixed-resolution spatial schedule, AIA maps
variable-level work onto customized RISC-V cores, and CoopMC co-designs the probability-generation and sampling pipeline.
Beyond hardware execution and sampling-pipeline optimization, \codename{} introduces an \ir{} that captures factorized computation and kernel-level dependencies, enabling compiler optimizations for irregular probabilistic graphical models. \codename{} achieves high sampling throughput while exposing multi-level parallel patterns
graph partitioning, dependency scheduling, and label-memory optimization to the compiler.


\section{Conclusion}
We presented \codename{}, an end-to-end compiler-driven accelerator framework for component-wise MCMC sampling over diverse discrete probabilistic graphical models.
Evaluation across structured and irregular energy-based models and causal-inference networks shows that \codename{} supports these workloads and achieves an average $6.1\times$ throughput improvement over the JAX-GPU baseline on irregular workloads. \codename{} is currently limited to discrete, component-wise MCMC and recompiles its static mapping when model topology changes.
Future work will extend this compiler--architecture interface to additional MCMC kernels and workloads with dynamically changing topology.
These results establish compiler-guided static mapping as a practical path toward programmable probabilistic-sampling accelerators.


\bibliography{refs}

\end{document}